# Active noise cancellation in ultra-low field MRI: distinct strategies for different channels

**Authors:** Jiali He[1], Sheng Shen[2,3], Jiamin Wu[4], Xiaohan Kong[5], Yamei Dai[1], Liang Tan[6], Zheng Xu[1*]

**Affiliations:**

[1]School of Electrical Engineering, Chongqing University, Chongqing 400044, China

[2]Athinoula A. Martinos Center for Biomedical Imaging, Massachusetts General Hospital. Charlestown, MA, United States of America (the)

[3]Harvard Medical School. Boston, MA, United States of America (the)

[4]Shenzhen Academy of Aerospace Technology, Shenzhen 518057, China.

[5]Graduate School of Engineering, Kyoto University, Kyoto 615-8530, Japan.

[6]Center of Critical Care Medicine, Southwest Hospital, Third Military Medical University, Chongqing, 400038, China

*Corresponding author. Email: xuzheng@cqu.edu.cn

**One Sentence Summary:** This study reveals that EMI coupling differs across channels in ultra-low field MRI systems. It further introduces an active cancellation strategy that improves noise suppression and channel consistency in open environments.

**Abstract:** Ultra-low field magnetic resonance imaging (ULF-MRI) systems operating in open environments are highly susceptible to composite electromagnetic interference (EMI). Different imaging channels respond non-uniformly to EMI owing to their distinct coupling characteristics. Here, we investigate channel-specific interference pathways in a permanent-magnet–based low-field MRI system and show that saddle coils are intrinsically more vulnerable to transverse EMI components than solenoidal coils. To mitigate these heterogeneous coupling effects, we propose a dual-stage suppression strategy that combines front-end spatial-domain inverse field reconstruction with back-end channel-adaptive active noise cancellation. Experiments demonstrate that this approach suppresses EMI by more than 80%, substantially improves inter-channel signal-to-noise ratio (SNR) consistency, and enhances the fused-image SNR by ~24%. These findings elucidate the channel-dependent nature of EMI coupling and establish targeted mitigation strategies, providing both a theoretical basis and practical guidance for noise suppression in future array-coil ULF-MRI systems.

**Main Text:**

## INTRODUCTION

Ultra-low field magnetic resonance imaging (ULF-MRI, magnetic field strength < 0.1 T) has attracted growing attention in recent years *(1–7)*, primarily due to its potential to enable low-cost, lightweight, and highly integrated imaging systems *(8–19)*. In contrast to conventional high-field MRI systems that rely on bulky superconducting magnets and liquid helium cooling, ULF-MRI can operate without cryogens by employing open permanent magnet architectures, thereby significantly reducing system complexity and operational costs. More importantly, its inherent electromagnetic safety and low power consumption make it particularly well-suited for in-vehicle imaging, bedside examinations, and essential medical services in resource-limited settings *(4)*.

Despite its notable engineering and application advantages, the clinical translation of ULF-MRI is still constrained by a fundamental technical bottleneck: the challenge of extremely low signal-to-noise ratio (SNR) *(3)*. The primary cause lies in the substantially weakened MRI signal strength at ultra-low magnetic fields, resulting in imaging SNRs far inferior to those of high-field systems. Crucially, ULF-MRI systems are often deployed in open or mobile environments, where shielding against external electromagnetic interference (EMI) is difficult to achieve. Ambient radiofrequency noise, power-line harmonics, and wireless communication signals are pervasive in such settings and can be readily picked up by the receive coils. These unwanted spectral components readily mix into the already weak MRI signal, leading to further degradation of SNR, compromised image quality, and potentially severe artifacts or distortions.

At present, mainstream approaches for EMI suppression are primarily based on passive shielding techniques *(20)*, with typical implementations including Faraday cages and portable shielding enclosures. While these methods can reduce interference to some extent, their limitations are becoming increasingly apparent. First, shielding structures substantially increase the overall size and weight of the system, thereby constraining portability and deployment flexibility. Second, they fail to effectively isolate EMI originating from internal electronic components, such as RF amplifiers, gradient power supplies, and digital control units. Moreover, enclosed shielding cabins may induce psychological discomfort in patients, an issue that is particularly pronounced in emergency and pediatric applications.

To address these challenges, researchers have proposed various active suppression strategies *(21–26)*. These approaches typically leverage the spatial and temporal characteristics of EMI signals, using sensor arrays to acquire interference waveforms in real time. Transfer function modeling is then applied to estimate the interference, followed by post-processing techniques for noise cancellation. Concurrently, deep learning has emerged as a new paradigm for EMI suppression *(1, 9, 27–29)*. Instead of relying on conventional channel modeling, this data-driven approach directly learns complex interference patterns from training datasets. Although both post-processing noise cancellation and deep learning significantly improve EMI mitigation, they still face key limitations, including high data acquisition demands, poor cross-device generalization, and limited physical interpretability. Furthermore, most existing studies focus on single-channel MRI, while the interference mechanisms in multi-channel systems remain insufficiently characterized.

To further improve imaging quality under ultra-low field conditions, multi-channel receive arrays have emerged as a key direction in ULF-MRI system development *(30–35)*. By leveraging parallel imaging techniques, such arrays markedly enhance acquisition efficiency *(36–39)*, a principle well established in high-field MRI *(40–44)* and now extending into low-field applications. However, several fundamental interference-related challenges persist in ULF-MRI systems. First, most existing studies focus primarily on the final suppression outcomes, with limited attention paid to the propagation and spatial distribution of EMI within the metallic shielding cavity. In confined spaces, complex electromagnetic behaviors such as reflection, standing waves, and coupling lead to interference evolution patterns that have not been systematically modeled, leaving their physical mechanisms largely unexplored. This challenge affects not only multi-channel configurations but also the broader ULF-MRI field. Second, MRI receive coils oriented in different directions exhibit varying sensitivity to external interference. For example, compared to commonly used solenoidal coils, saddle-shaped coils are more prone to capturing environmental EMI in the transverse direction, which significantly raises the noise floor during imaging and leads to severe SNR imbalance *(25, 45)*. This heightened structural sensitivity to interference has become a major technical bottleneck limiting the performance of multi-channel ULF-MRI systems.

In this study, we investigate the coupling pathways of EMI within the metallic enclosures of unshielded ULF-MRI systems, revealing a pronounced direction-dependent behavior. Theoretical modeling and numerical simulations show that, in a single-sided-open rectangular cavity, transverse EMI components dominate over longitudinal ones, following a "transverse-dominant, longitudinal-weak" distribution that decays gradually from the opening inward. Experimental measurements confirm these trends, demonstrating the universality and stability of this spatial pattern. Guided by these insights, we propose a spatial-domain active suppression method based on wave-interference principles. By constructing a cancellation system capable of modulating both phase and amplitude, the approach achieves real-time spatial cancellation of interference within the imaging region, complementing and extending conventional post-processing strategies. When applied to saddle coils, this method substantially improves imaging SNR. Comparative experiments under diverse interference conditions further demonstrate its robustness and adaptability. Beyond suppressing EMI, this work highlights the direction-dependent coupling between coil geometries and environmental noise, providing a physically grounded framework and an initial engineering pathway toward high-performance multi-channel ULF-MRI.

**RESULTS**

To address the SNR limitations imposed by composite EMI in ULF-MRI, we first examined the underlying coupling mechanisms. Our analysis revealed a strong directional effect of metal cavities, with transverse EMI components ($H_y$) dominating within the imaging region and exhibiting distinct spatial distribution patterns. Motivated by the pronounced noise contamination observed in saddle-coil channels, we developed an active cancellation strategy grounded in wave-interference principles. Specifically, we designed a phase- and amplitude-tunable cancellation system capable of dynamically matching the interference field, thereby enabling front-end suppression of noise before signal acquisition.

We then implemented and experimentally validated this approach using a dual-channel ULF-MRI system in an unshielded environment. The results demonstrate that the combined strategy improves the SNR of saddle coils and reduces inter-channel imbalance, providing a practical route toward achieving higher-quality ULF-MRI in open and interference-prone settings.

**EMI Field Distribution Inside the Imaging Cavity**

This study was conducted on a custom-built 0.05 T ULF-MRI platform (Fig. 1A). The system incorporates a pair of noise-cancellation coils, wound more densely on the outer side and more sparsely on the inner side, to generate an inverted EMI field. Two types of noise detection coils were deployed: one to provide signals for front-end spatial-domain cancellation, and another to supply data for post-processing cancellation. The platform integrates compact imaging electronics and employs orthogonal solenoidal and saddle coils as the quadrature receive unit. An implementation in a real imaging scenario is shown in Fig. 1B, with the internal structure detailed in Fig. S1.

In an unshielded environment, in situ measurements revealed that the EMI field inside the imaging cavity exhibits pronounced directional coupling characteristics (The measurement platform is shown in Fig. S2.). We measured the EMI across the entire cavity of the system. We focused our analysis on the critical imaging target region—a sphere with a radius of 100 mm centered at the cavity origin. The results indicate that, within this region, the transverse magnetic field component constitutes the dominant source of electromagnetic noise. Specifically, the interference field was decomposed into a longitudinal component ($Hx$, along the x-axis, primarily coupling to the solenoidal coil) and a transverse component ($Hy$, along the y-axis, primarily coupling to the saddle coil), while the z-axis component was excluded due to its alignment with the static magnetic field. Experimental measurements (Figs. 2A and 2C) indicate that the $Hx$ component exhibits minimal intensity at the cavity center (y = 0) but significantly increases toward both sides along the y-axis. Conversely, the $Hy$ component peaks at the center and gradually diminishes toward the periphery (Figs. 2B and 2C). Notably, within the core imaging region, the intensity of $Hy$ is markedly higher than that of $Hx$ (Fig. 2C). Furthermore, both components display a clear gradient attenuation along the cavity's depth (x-axis) direction (Figs. 2A–C).

To gain deeper insights into the spatial distribution of EMI within the metal cavity and its direction-dependent coupling with coil geometries, simulation analyses were conducted using a 2.23 MHz plane wave as the interference source (Fig. 3). The simulation aimed to characterize the distribution of different field components ($Hx$ and $Hy$) within the imaging region, the influence of the incident wave direction on coupling efficiency, and the non-propagating nature of the interference field within the cavity. The results (Fig. 3A) corroborate the experimental observations: $Hx$ forms a distribution with opposite directions and symmetrical amplitudes on both sides of the cavity, approaching zero at the center; meanwhile, the amplitude of $Hy$ is strongest in the central region and gradually weakens toward both sides of the y-axis, while maintaining a uniform direction throughout. Simulations further demonstrate that when the wave vector $\boldsymbol{k}$ is oriented perpendicularly toward the cavity opening (-x direction), the magnetic field coupling efficiency reaches its maximum (Figs. 3A, B). As the incident direction deviates from this axis—by rotating the electric or magnetic field vector—the coupling efficiency decreases, though the spatial distribution of the field remains consistent (Figs. 3A, B). Additionally, for any incident angle, both $Hx$ and $Hy$ components exhibit rapid decay along the x-axis inside the cavity, confirming their non-propagating field behavior (Fig. 3A).

Together, the measurements and simulations establish three key findings: (i) In unshielded conditions, EMI within the cavity exhibits strong directionality—$Hx$ is minimal at the center and increases laterally, whereas $Hy$ is maximal at the center and decreases toward the sides; both attenuate along the cavity depth. (ii) The transverse component $Hy$ constitutes the dominant noise source within the imaging region. (iii) The observed SNR imbalance between saddle and solenoidal coils originates not from coil geometry but from the intrinsic field distribution: $Hy$ is strong and aligned in a single direction at the center, while $Hx$ is weak and antisymmetric, leading to net flux cancellation.

**Active EMI Suppression Strategy**

Building on the finding that the transverse EMI component ($Hy$) within the imaging region is significantly stronger than the longitudinal component and constitutes the primary limitation on the SNR of the saddle coil channel, this study proposes and implements a dynamic active cancellation strategy based on wave interference principles. This strategy generates, in real-time, an inverse magnetic field that precisely matches the phase and amplitude of the $Hy$ interference but with opposite direction, achieving spatial cancellation and purification of the interference signals within the imaging area. Considering the saddle coil's high sensitivity to the $Hy$ component and the solenoidal coil's negligible response to $Hx$ interference, a differentiated noise reduction scheme is designed for each channel: a combined spatial-domain active suppression and post-processing method for the saddle coil, and conventional post-processing only for the solenoidal coil, thereby balancing noise suppression efficacy with system stability (Fig. 4).

For saddle-coil channels severely affected by $Hy$ interference, spatial-domain active noise control (ANC) is employed. A dedicated noise-detection coil captures the environmental $Hy$ interference in real time, and the signal is precisely modulated in phase and amplitude by high-speed analog circuits. The modulated signal then drives customized transverse cancellation coils symmetrically arranged on both sides of the cavity (Fig. 5A), generating a magnetic field in the target region that is precisely opposite to the original $Hy$ interference. After effectively canceling the transverse interference field, the low-noise MR signals acquired by the saddle coil are further processed by conventional post-processing ANC methods to suppress residual noise (24) (Fig. 4C) (Fig. S3). In contrast, the solenoidal channel, owing to its intrinsic resistance to $Hy$ interference, only requires conventional post-processing to obtain high-quality data (Fig. 4B).

To precisely match the gradient distribution characteristics of the $Hy$ interference field in the target region, a transverse cancellation coil was designed (Fig. 5A). This coil adopts a winding configuration with denser turns on the outer side and sparser turns on the inner side (Fig. 5B), ensuring that the generated cancellation field is predominantly oriented along the y-axis and exhibits a spatial gradient consistent with that of the interference field. Performance validation (Figs. 5B and S4) demonstrates that the coil's x-axis magnetic field component exhibits a symmetric bipolar distribution in the x–z plane, with opposite directions and magnitudes significantly smaller than the y-axis component. Consequently, the net magnetic flux induced in the solenoidal coil is zero, introducing no additional interference. The z-axis magnetic field component is an order of magnitude weaker than the y-axis component and is parallel to the static magnetic field (z-axis), thus exerting no significant influence on MRI signal acquisition. Simulation results (Fig. 5C) show that this coil design reduces the target-region interference field strength from 0.07–0.12 A/m before cancellation to 0–0.03 A/m after cancellation. From the perspective of the saddle coil's detected EMI flux, the transverse EMI is reduced by 84.2%, indicating a substantial noise suppression effect.

After channel-specific noise reduction, the SNRs of the two channels converged to comparable levels. Adaptive weighted image fusion (Fig. 4D) further enhanced overall image quality. Simulations demonstrated that: (i) the active cancellation strategy reduced $Hy$ in the saddle-coil region to less than 15.8% of its original level; (ii) the gradient-matched coil design effectively avoided interference with both the solenoidal channel and the static field; and (iii) the combined use of channel-specific active cancellation, post-processing, and adaptive fusion markedly improved the SNR of multi-channel ULF-MRI under interference-prone conditions.

**Experimental Validation**

Human brain imaging (Fig. 6) and water phantom (Fig. S5) experiments were conducted on a self-developed unshielded ULF-MRI system to verify the feasibility of the active interference suppression strategy. The system performance was evaluated under three conditions: conventional post-processing, spatial-domain active suppression, and their combined approach. A dual-channel imaging platform (saddle coil + solenoid coil) was established in an unshielded environment. Experimental results showed that the solenoid channel's SNR in the original state (18.21 dB) was three times that of the saddle coil channel (9.18 dB) (Fig. 6A-B). This discrepancy arises from the directional coupling effect of EMI within the metal cavity— the saddle coil is over three times more sensitive to transverse interference than the solenoid coil (Fig. 7A), and this disparity amplifies with increasing environmental interference (Fig. S6).

Using single noise reduction alone (either post-processing active noise reduction or spatial active noise reduction), the SNR of the saddle coil channel improved only to 18 dB (Fig. 6A), comparable to the original solenoid channel SNR, which is insufficient for high-quality imaging. In contrast, the combined noise reduction strategy achieves a breakthrough through a hierarchical suppression: the spatial-domain active cancellation device suppresses 76% of transverse interference at the signal acquisition front-end (reducing the maximum from 300.32 to 72.19; Fig. 7C), lowering the saddle coil channel background noise to nearly the same level as the solenoid channel (Fig. 7). Subsequently, the post-processing active noise reduction algorithm (or other conventional post-processing methods) further eliminates residual noise. This strategy raises the saddle coil channel SNR to 26.87 dB (Fig. 6A), comparable to the post-processed solenoid channel level (Fig. 6B). Adaptive weighted fusion of the two-channel images results in a 24% improvement in overall system SNR (Fig. 6C). Noise analysis (Fig. 7) reveals the optimization mechanism of the combined strategy: although conventional post-processing reduces overall noise (Fig. 7B), residual interference in the saddle coil channel remains twice that of the solenoid; spatial-domain active cancellation balances the dual-channel noise levels at the source (Fig. 7C), laying the foundation for fusion reconstruction.

This dual-interference suppression scheme not only effectively attenuates transverse background interference at its origin but also enhances SNR quality and channel balance during image reconstruction, demonstrating significant potential to enable high-quality multi-channel ULF-MRI imaging in unshielded natural environments.

**DISCUSSION**

This study addresses an issue in ULF-MRI systems that remains insufficiently explored: the channel-specific coupling mechanisms of EMI in open, unshielded environments. Such variations can markedly impair the quality and consistency of multichannel imaging. Through theoretical analysis and numerical modeling, we demonstrate that the metallic cavity boundary modulates the distribution of interference fields, inducing direction-selective, non-propagating modes. This asymmetry leads to substantial differences in interference intensity across channels. Building on this insight, we developed a dual-stage denoising strategy that combines front-end reverse magnetic field generation with backend signal processing. This approach actively cancels the dominant interference components and significantly improves both inter-channel SNR consistency and image fusion quality. Unlike conventional methods that rely solely on post-processing, our strategy emphasizes spatial-domain physical modulation of the interference field prior to signal acquisition. By treating EMI as a modelable and controllable vector field, we offer channel-specific solutions for interference mitigation in low-field MRI systems, establishing both theoretical and technical foundations for denoising in future array coil architectures.

We found that the variability in channel-specific EMI response originates primarily from the directional coupling between external noise fields and the magnet structure. Specifically, when EMI enters through the cavity opening, the metal walls constrain the electric field, disrupting the original wave pattern and inducing a non-propagating, decaying field dominated by the $Hy$ component in the imaging volume. This produces transverse much stronger than longitudinal magnetic fields. As saddle coils are highly sensitive to transverse components, their SNR is markedly degraded*(25, 45)*, while solenoid coils—primarily responsive to longitudinal components—remain relatively unaffected. These observations (Fig. 2–3) underscore that the response of a coil to interference is not solely determined by its geometry, but rather by the combined effect of system geometry, boundary conductance, and the incident EMI field, providing a critical understanding of inter-channel heterogeneity in multichannel systems.

Effective field cancellation depends on precise phase matching between the interference and compensating fields, which arises from the phase dynamics of magnetic induction. Since the induced voltage in the acquisition coil leads magnetic flux variation by 90°, the circuit must impose an additional 270° phase delay to generate a 180° out-of-phase cancellation field. Given spatial propagation contributes only a negligible phase delay (<3°), phase control is concentrated within the circuit to improve system stability and signal controllability. Impedance matching is employed to maintain the compensation coil as a purely resistive load, ensuring phase alignment between drive signals and output current. This mechanism underscores that suppressing low-frequency transverse interference is fundamentally a vector field modulation process governed by coordinated phase-amplitude control—its precision directly impacting EMI suppression effectiveness.

The proposed strategy of "vector-field interference modeling and active inverse response" demonstrates strong scalability. Unlike conventional passive filtering, this approach models electromagnetic coupling within the cavity as a reversible control problem, thereby establishing a novel front-end active EMI suppression framework for ULF-MRI systems. Beyond its application to saddle coil configurations, it also provides a theoretical basis for directional noise control in other parallel imaging systems. Our analysis shows that in multichannel arrays, the primary source of SNR degradation is the non-uniform distribution of transverse EMI, arising from variations in coil location and sensitivity orientation. By introducing an inverse transverse interference field, the intensity within the imaging volume can be reduced to the level of the longitudinal component. Thereby creates a cleaner electromagnetic environment and offers a practical solution to the long-standing imbalance between imaging channels.

Nevertheless, several limitations remain, as the current compensation coil design is still at a preliminary stage. While it effectively suppresses the dominant $H_y$ interference, imperfect gradient matching leads to incomplete cancellation and residual interference in localized regions (Fig. 5E). To enhance system adaptability and robustness, a generalized coil topology was used in this study. Future work may involve tailoring coil geometries to the spatial distribution of the interference field in the target region, improving both uniformity and local precision.

Further developments are needed to meet the remaining challenges. In terms of active EMI suppression, future work should improve the spatial fitting accuracy and response speed of the inverse field, possibly incorporating AI-based field matching and structural optimization algorithms to enhance adaptability to complex interference environments. Inverse problem solving and intelligent control strategies may enable higher-resolution and more stable EMI cancellation. On the system integration front, specialized drive circuits with low noise and high responsiveness could improve overall performance and practicality. The current setup has shown strong robustness in unshielded environments, establishing a solid foundation for future clinical adaptation. Ongoing studies will focus on imaging stability, channel consistency, and patient-friendly user interfaces. Initial clinical trials using real patient data will evaluate its effectiveness in lesion detection and anatomical imaging, paving the way toward practical implementation of ULF-MRI in open, mobile environments.

In conclusion, this work presents a comprehensive framework for active EMI suppression in low-field MRI systems, encompassing theoretical modeling, system design, and experimental validation. By enabling spatial-domain control of directional interference at the source, the proposed method significantly improves dual-channel imaging quality and fusion consistency under unshielded conditions. This strategy lays a scalable, forward-looking foundation for the development of low-power, portable, multichannel ULF-MRI systems.

## MATERIALS AND METHODS

### ULF-MRI System and Experimental Environment

The experimental platform was a custom-built, portable, unshielded 50 mT MRI scanner. The core magnet was a samarium–cobalt (SmCo) permanent magnet that generated a 0.05 T static field along the z-axis, with field inhomogeneity below 70 ppm within a 200 mm spherical imaging volume (DSV). The imaging cavity was constructed from low-carbon steel (880 × 590 × 480 mm³), forming a magnetic yoke on four sides, a copper mesh boundary at the rear, and a single open side.

For portability, all electronic modules—gradient power amplifier, RF power amplifier, spectrometer, and RF switching unit—were enclosed in an aluminum housing mounted on a mobile frame that also supported the magnet. The preamplifier was positioned separately to minimize noise. The system operated at a central frequency of 2.2 MHz. Gradient coil efficiencies were 26.7 mT/m/100 A along the X and Y axes and 43.5 mT/m/100 A along the Z axis. The RF coil was tuned to the Larmor frequency (2.2 MHz) with a 100 kHz bandwidth. Signal acquisition was performed using an MR solution EVO spectrometer (MR Solutions, UK) with 8 receive channels and 1 transmit channel. The linear gradient amplifier was capable of driving coils with inductances below 500 µH.

All experiments were conducted in a standard building environment without electromagnetic shielding. Sources of interference included other MRI scanners, electric fans, LED lights, computers, and mobile phones. Unless otherwise specified, all measurements were acquired under these ambient conditions.

During imaging, spatial-domain detection coils continuously monitored transverse EMI components and drove cancellation coils in real time through an analog inversion circuit, thereby suppressing interference within the cavity. EMI signals from multiple detection coils were simultaneously recorded along with MR signals for post-processing. All imaging experiments employed a GRE sequence with three signal averages, resulting in a scan time of approximately 6 minutes per acquisition.

### EMI measurement and simulation model

To characterize spatial variations of magnetic interference, we developed a custom EMI measurement system based on a three-axis positioning stage (Fig. S2). Because environmental noise is inherently time-varying, the system incorporated a five-point synchronized sampling array, with adjacent pickup coils spaced 15 cm apart—the minimum distance required to avoid mutual inductive coupling. Detection coils were arranged along both the X and Y directions on the top surface of the imaging system. Field measurements inside the cavity were normalized to simultaneously acquired external probe data to compensate for temporal noise fluctuations. A total of 225 measurement points within a 60 × 50 × 20 cm³ volume were uniformly sampled in both transverse and longitudinal directions, with each point measured five times for averaging.

Numerical simulations were performed using Ansys HFSS (Fig. 3). The imaging cavity was modeled as a rectangular metallic enclosure (880 × 590 × 480 mm³, wall thickness 50 mm) composed of low-carbon steel. All faces were closed except for a single open side, forming a magnetic yoke boundary, with the rear wall represented as a copper mesh. An external noise source located at x = -λ generated a 2.23 MHz plane wave with the electric field oriented along z and the magnetic field along y, yielding a Poynting vector directed along -x, perpendicular to the cavity opening. To examine direction-dependent coupling, the incident wave vector was rotated in 15° increments about both the *H* and *E* axes.

## Spatial-Domain Active Noise Cancellation System

To suppress the dominant transverse interference ($H_y$) identified in the imaging region, a spatial-domain ANC system was implemented. The principle is to generate a compensatory magnetic field $\boldsymbol{B_C(r, t)}$ that superimposes with the interference field $\boldsymbol{B_{EMI}(r, t)}$, such that $\boldsymbol{B_{EMI}(r, t)} + \boldsymbol{B_C(r, t)} = \boldsymbol{0}$, thereby reducing noise and improving SNR in MRI measurements. Because the operating frequency (2.23 MHz; wavelength ≈ 134.5 m) is much larger than the magnet dimensions (<2 m), the system can be modeled under the quasi-static approximation, where propagation delays are negligible and the spatial distribution of the magnetic field can be decoupled from its temporal dynamics. This allows real-time synthesis of counteracting fields.

The ANC circuit adopted an analog architecture to enable rapid phase and amplitude control with minimal added noise. The system comprises a low-noise preamplifier, in-phase and inverting amplifiers, a tunable phase-compensation module, and a driver coil unit. The detection coil output was impedance-matched to minimize reflection and loss, then amplified to preserve SNR. A two-stage operational amplifier with adjustable feedback provided both gain tuning and a 270° phase delay. After phase compensation, the processed signal drove the cancellation coil to generate the inverted magnetic field. This architecture is tunable and adaptable to different interference conditions, providing flexibility for integration into multichannel ULF-MRI systems.

## List of Supplementary Materials

Fig. S1 to S6 for multiple supplementary figures

## References and Notes

1. Y. Zhao, Y. Ding, V. Lau, C. Man, S. Su, L. Xiao, A. T. L. Leong, E. X. Wu, Whole-body magnetic resonance imaging at 0.05 Tesla. *Science* **384**, eadm7168 (2024).

2. U. C. Anazodo, S. D. Plessis, Imaging without barriers. *Science* **384**, 623–624 (2024).

3. T. C. Arnold, C. W. Freeman, B. Litt, J. M. Stein, Low-field MRI: Clinical promise and challenges. *Journal of Magnetic Resonance Imaging* **57**, 25–44 (2023).

4. S. Murali, H. Ding, F. Adedeji, C. Qin, J. Obungoloch, I. Asllani, U. Anazodo, N. A. B. Ntusi, R. Mammen, T. Niendorf, S. Adeleke, Bringing MRI to low- and middle-income countries: Directions, challenges and potential solutions. *NMR in Biomedicine* **37**, e4992 (2024).

5. C. N. DesRoche, A. P. Johnson, E. B. Hore, E. Innes, I. Silver, D. Tampieri, B. Y. M. Kwan, J. O. Jimenez, J. G. Boyd, O. Islam, Feasibility and Cost Analysis of Portable MRI Implementation in a Remote Setting in Canada. *Canadian Journal of Neurological Sciences* **51**, 387–396 (2024).

6. A. Altaf, M. Shakir, H. A. Irshad, S. Atif, U. Kumari, O. Islam, W. T. Kimberly, E. Knopp, C. Truwit, K. Siddiqui, S. A. Enam, Applications, limitations and advancements of ultra-low-field magnetic resonance imaging: A scoping review. *Surg Neurol Int* **15**, 218 (2024).

7. M. Sarracanie, N. Salameh, Low-Field MRI: How Low Can We Go? A Fresh View on an Old Debate. *Front. Phys.* **8** (2020), doi:10.3389/fphy.2020.00172.

8. M. H. Mazurek, B. A. Cahn, M. M. Yuen, A. M. Prabhat, I. R. Chavva, J. T. Shah, A. L. Crawford, E. B. Welch, J. Rothberg, L. Sacolick, M. Poole, C. Wira, C. C. Matouk, A. Ward, N. Timario, A. Leasure, R. Beekman, T. J. Peng, J. Witsch, J. P. Antonios, G. J. Falcone, K. T. Gobeske, N. Petersen, J. Schindler, L. Sansing, E. J. Gilmore, D. Y. Hwang, J. A. Kim, A. Malhotra, G. Sze, M. S. Rosen, W. T. Kimberly, K. N. Sheth, Portable, bedside, low-field magnetic resonance imaging for evaluation of intracerebral hemorrhage. *Nat Commun* **12**, 5119 (2021).

9. Y. Liu, A. T. L. Leong, Y. Zhao, L. Xiao, H. K. F. Mak, A. C. O. Tsang, G. K. K. Lau, G. K. K. Leung, E. X. Wu, A low-cost and shielding-free ultra-low-field brain MRI scanner. *Nat Commun* **12**, 7238 (2021).

10. J. Zhang, C. Pu, J. Wang, S. Liu, R. Lu, H. Yi, in 2021 IEEE 15th International Conference on Electronic Measurement & Instruments (ICEMI), (2021), pp. 32–36.

11. A. M. Prabhat, A. L. Crawford, M. H. Mazurek, M. M. Yuen, I. R. Chavva, A. Ward, W. V. Hofmann, N. Timario, S. R. Qualls, J. Helland, C. Wira, G. Sze, M. S. Rosen, W. T. Kimberly, K. N. Sheth, Methodology for Low-Field, Portable Magnetic Resonance Neuroimaging at the Bedside. *Front. Neurol.* **12** (2021), doi:10.3389/fneur.2021.760321.

12. C. Z. Cooley, P. C. McDaniel, J. P. Stockmann, S. A. Srinivas, S. F. Cauley, M. Śliwiak, C. R. Sappo, C. F. Vaughn, B. Guerin, M. S. Rosen, M. H. Lev, L. L. Wald, A portable scanner for magnetic resonance imaging of the brain. *Nat Biomed Eng* **5**, 229–239 (2021).

13. M. Sarracanie, C. D. LaPierre, N. Salameh, D. E. J. Waddington, T. Witzel, M. S. Rosen, Low-Cost High-Performance MRI. *Sci Rep* **5**, 15177 (2015).

14. C. Z. Cooley, J. P. Stockmann, B. D. Armstrong, M. Sarracanie, M. H. Lev, M. S. Rosen, L. L. Wald, Two-dimensional imaging in a lightweight portable MRI scanner without gradient coils. *Magnetic Resonance in Medicine* **73**, 872–883 (2015).

15. C. Z. Cooley, M. W. Haskell, S. F. Cauley, C. Sappo, C. D. Lapierre, C. G. Ha, J. P. Stockmann, L. L. Wald, Design of Sparse Halbach Magnet Arrays for Portable MRI Using a Genetic Algorithm. *IEEE Transactions on Magnetics* **54**, 1–12 (2018).

16. S. Lother, S. J. Schiff, T. Neuberger, P. M. Jakob, F. Fidler, Design of a mobile, homogeneous, and efficient electromagnet with a large field of view for neonatal low-field MRI. *MAGMA* **29**, 691–698 (2016).

17. L. M. Broche, P. J. Ross, G. R. Davies, M.-J. MacLeod, D. J. Lurie, A whole-body Fast Field-Cycling scanner for clinical molecular imaging studies. *Sci Rep* **9**, 10402 (2019).

18. T. O'Reilly, A. Webb, Deconstructing and reconstructing MRI hardware. *Journal of Magnetic Resonance* **306**, 134–138 (2019).

19. Y. He, W. He, L. Tan, F. Chen, F. Meng, H. Feng, Z. Xu, Use of 2.1 MHz MRI scanner for brain imaging and its preliminary results in stroke. *Journal of Magnetic Resonance* **319**, 106829 (2020).

20. T. Guallart-Naval, J. M. Algarín, R. Pellicer-Guridi, F. Galve, Y. Vives-Gilabert, R. Bosch, E. Pallás, J. M. González, J. P. Rigla, P. Martínez, F. J. Lloris, J. Borreguero, Á. Marcos-Perucho, V. Negnevitsky, L. Martí-Bonmatí, A. Ríos, J. M. Benlloch, J. Alonso, Portable magnetic resonance imaging of patients indoors, outdoors and at home. *Sci Rep* **12**, 1–11 (2022).

21. S. Hushek, J. Saunders, J. Schellenberg, Automatic Noise Cancellation for Unshielded Mr Systems (2008) (available at https://patentscope.wipo.int/search/en/detail.jsf?docId=WO2008022441).

22. T. Rearick, G. L. Charvat, M. S. Rosen, J. M. Rothberg, Noise suppression methods and apparatus (2017) (available at https://patents.google.com/patent/US9797971B2/en).

23. S. A. Srinivas, S. F. Cauley, J. P. Stockmann, C. R. Sappo, C. E. Vaughn, L. L. Wald, W. A. Grissom, C. Z. Cooley, External Dynamic InTerference Estimation and Removal (EDITER) for low field MRI. *Magnetic Resonance in Medicine* **87**, 614–628 (2022).

24. L. Yang, W. He, Y. He, J. Wu, S. Shen, Z. Xu, Active EMI Suppression System for a 50 mT Unshielded Portable MRI Scanner. *IEEE Transactions on Biomedical Engineering* **69**, 3415–3426 (2022).

25. Y. Liu, L. Xiao, M. Lyu, R. Zhu, Eliminating electromagnetic interference for RF shielding-free MRI via k-space convolution: Insights from MR parallel imaging advances. *Journal of Magnetic Resonance* **369**, 107808 (2024).

26. S. Biber, S. Kannengiesser, J. Nistler, M. Braun, S. Blaess, M. Gebhardt, D. Grodzki, D. Ritter, G. Seegerer, M. Vester, R. Schneider, Design and operation of a whole-body MRI scanner without RF shielding. *Magnetic Resonance in Med* **93**, 1842–1855 (2025).

27. J. Su, R. Pellicer-Guridi, T. Edwards, M. Fuentes, M. S. Rosen, V. Vegh, D. Reutens, A CNN Based Software Gradiometer for Electromagnetic Background Noise Reduction in Low Field MRI Applications. *IEEE Transactions on Medical Imaging* **41**, 1007–1016 (2022).

28. Y. Zhao, L. Xiao, J. Hu, E. X. Wu, Robust EMI elimination for RF shielding-free MRI through deep learning direct MR signal prediction. *Magnetic Resonance in Medicine* **92**, 112–127 (2024).

29. Y. Zhao, L. Xiao, Y. Liu, A. T. Leong, E. X. Wu, Electromagnetic interference elimination via active sensing and deep learning prediction for radiofrequency shielding-free MRI. *NMR in Biomedicine* **37**, e4956 (2024).

30. S. Shen, Y. Zhang, X. Kong, L. Yang, M. S. Rosen, Z. Xu, An Exploration of a Phased-Array RF Coil for Very Low-Field Brain MRI. *IEEE Sensors J.* **24**, 2905–2914 (2024).

31. L. Yang, W. He, S. Shen, Y. He, J. Wu, Z. Xu, An Exploration of Parallel Imaging System for Very-Low-Field (50 mT) MRI Scanner. *IEEE Transactions on Instrumentation and Measurement* **74**, 1–13 (2025).

32. Q. Chen, Y. Xu, Y. Chang, X. Yang, Design and Demonstration of Four-Channel Received Coil Arrays for Vertical-Field MRI. *Appl Magn Reson* **48**, 501–515 (2017).

33. X. He, R. Yuan, B. K. Li, Y. Hou, The Design of an Open MRI 4-Channel Receive-Only Phased Array Knee Coil. *Appl Magn Reson* **47**, 499–510 (2016).

34. H. Fujita, T. Zheng, X. Yang, M. J. Finnerty, S. Handa, RF Surface Receive Array Coils: The Art of an LC Circuit. *Journal of Magnetic Resonance Imaging* **38**, 12–25 (2013).

35. H. Fujita, New horizons in MR technology: RF coil designs and trends. *Magn Reson Med Sci* **6**, 29–42 (2007).

36. F. Muñoz, Y. Lim, S. X. Cui, H. Stark, K. S. Nayak, Evaluation of a novel 8-channel RX coil for speech production MRI at 0.55 T. *Magn Reson Mater Phy* **36**, 419–426 (2023).

37. B. Li, R. Xie, Z. Sun, X. Shao, Y. Lian, H. Guo, R. You, Z. You, X. Zhao, Nonlinear metamaterials enhanced surface coil array for parallel magnetic resonance imaging. *Nat Commun* **15**, 7949 (2024).

38. M. C. Restivo, R. Ramasawmy, W. P. Bandettini, D. A. Herzka, A. E. Campbell-Washburn, Efficient spiral in-out and EPI balanced steady-state free precession cine imaging using a high-performance 0.55T MRI. *Magnetic Resonance in Medicine* **84**, 2364–2375 (2020).

39. H. Wang, W. Feng, X. Ren, Q. Tao, L. Rong, Y. P. Du, H. Dong, Acquisition Acceleration of Ultra-Low Field MRI With Parallel Imaging and Compressed Sensing in Microtesla Fields. *IEEE Trans. Biomed. Eng.* **72**, 655–663 (2025).

40. X. Yan, C. Ma, L. Shi, Y. Zhuo, X. J. Zhou, L. Wei, R. Xue, Optimization of an 8-Channel Loop-Array Coil for a 7 T MRI System with the Guidance of a Co-Simulation Approach. *Appl Magn Reson* **45**, 437–449 (2014).

41. J. Cohen-Adad, A. Mareyam, B. Keil, J. R. Polimeni, L. L. Wald, 32-Channel RF coil optimized for brain and cervical spinal cord at 3 T. *Magnetic Resonance in Medicine* **66**, 1198–1208 (2011).

42. C. J. Hardy, R. O. Giaquinto, J. E. Piel, K. W. Rohling AAS, L. Marinelli, D. J. Blezek, E. W. Fiveland, R. D. Darrow, T. K. F. Foo, 128-channel body MRI with a flexible high-density receiver-coil array. *Journal of Magnetic Resonance Imaging* **28**, 1219–1225 (2008).

43. J. R. Corea, A. M. Flynn, B. Lechêne, G. Scott, G. D. Reed, P. J. Shin, M. Lustig, A. C. Arias, Screen-printed flexible MRI receive coils. *Nat Commun* **7**, 10839 (2016).

44. B. Zhang, D. K. Sodickson, M. A. Cloos, A high-impedance detector-array glove for magnetic resonance imaging of the hand. *Nat Biomed Eng* **2**, 570–577 (2018).

45. Y. Huang, S. Qu, Y. Xie, H. Wang, X. Zhang, X. Zhang, Inter-Channel Correlation-Based EMI Noise Removal (ICER) for Shielding-Free Low-Field MRI. *IEEE Transactions on Biomedical Engineering* **72**, 2095–2104 (2025).

**Acknowledgments:** The author would like to thank Lei Yang, Yuxiang Zhang, and Cai Wan for their guidance on the experiment.

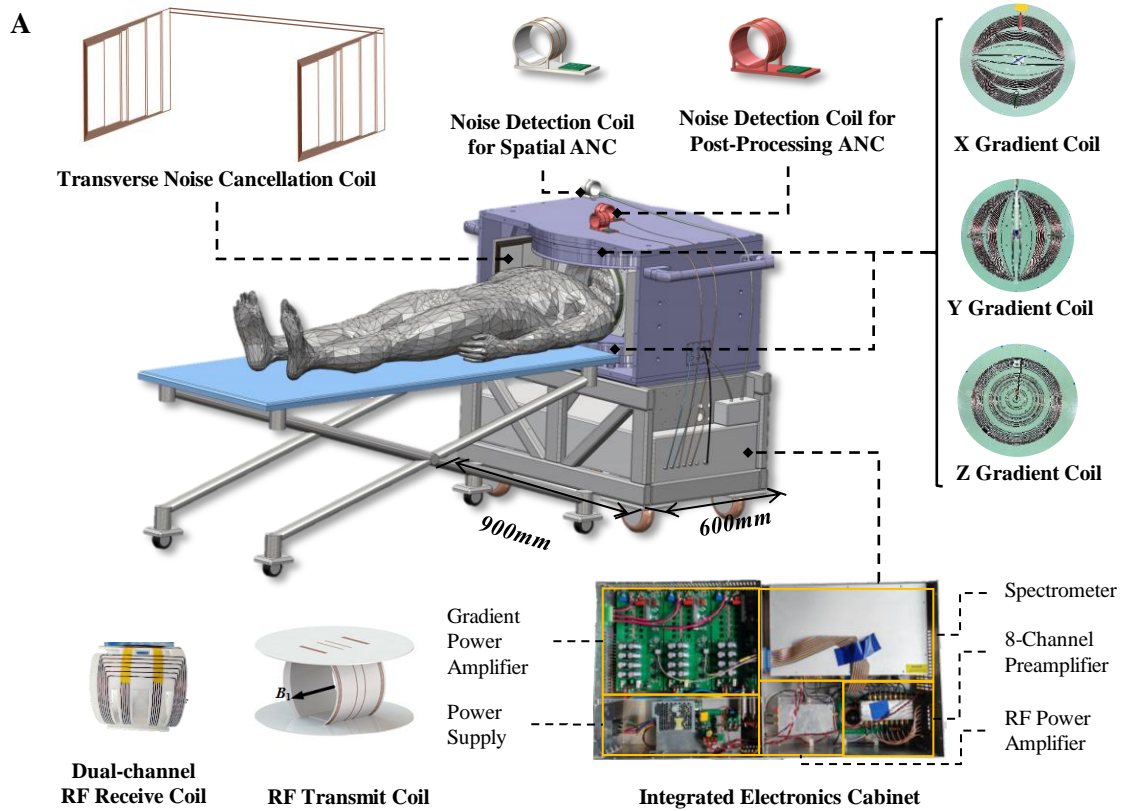

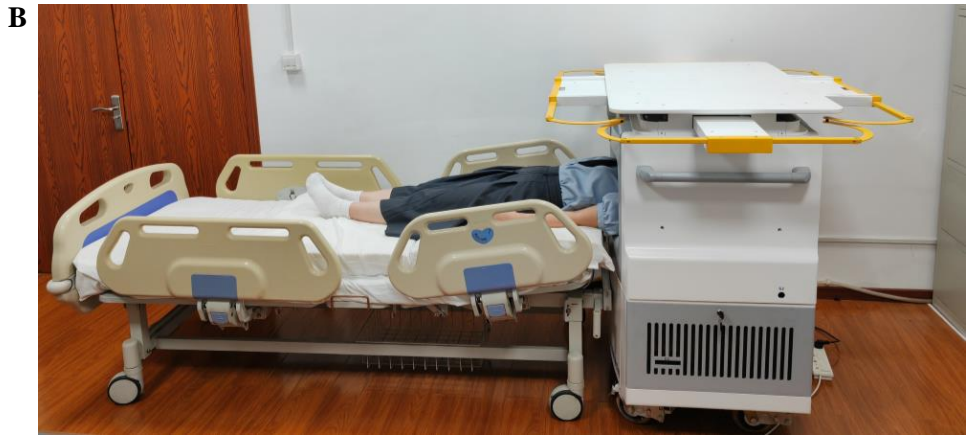

**Fig. 1. Portable 0.05 T ULF-MRI system. (A)** Schematic of the compact scanner (footprint <1 m²), showing the RF transmit coil, orthogonal saddle and solenoidal receive coils, and three-axis gradient coils. A pair of noise-cancellation coils with a dense–sparse winding pattern is mounted on both sides of the cavity to generate an inverse interference field. Two types of noise-detection coils are incorporated: one for spatial-domain active suppression and another for post-processing suppression. Integrated electronics include gradient amplifiers, a spectrometer, an 8-channel preamplifier array, and an RF power amplifier. **(B)** Photograph of the experimental platform used for human imaging.

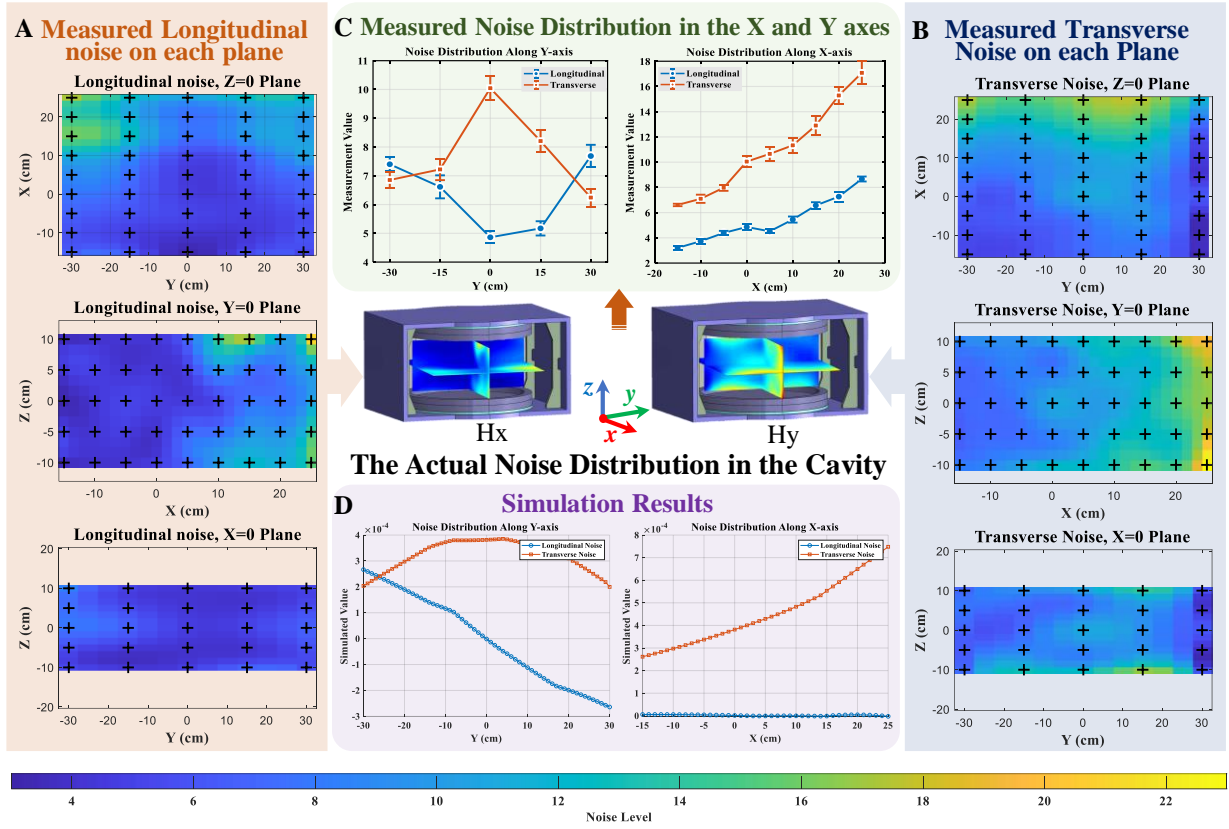

**Fig. 2. Measured and simulated EMI in the imaging cavity.** (A) Spatial distribution of longitudinal magnetic (*Hx*) interference fields on three representative planes (Z=0, Y=0, X=0). (B) Spatial distribution of transverse magnetic (*Hy*) interference fields on the same planes. (C) Profiles of measured *Hx* and *Hy* fields along the Y- and X-axes at the cavity center. (D) Simulated *Hx* and *Hy* field distributions along the Y- and X-axes at the cavity center for comparison with experimental results.

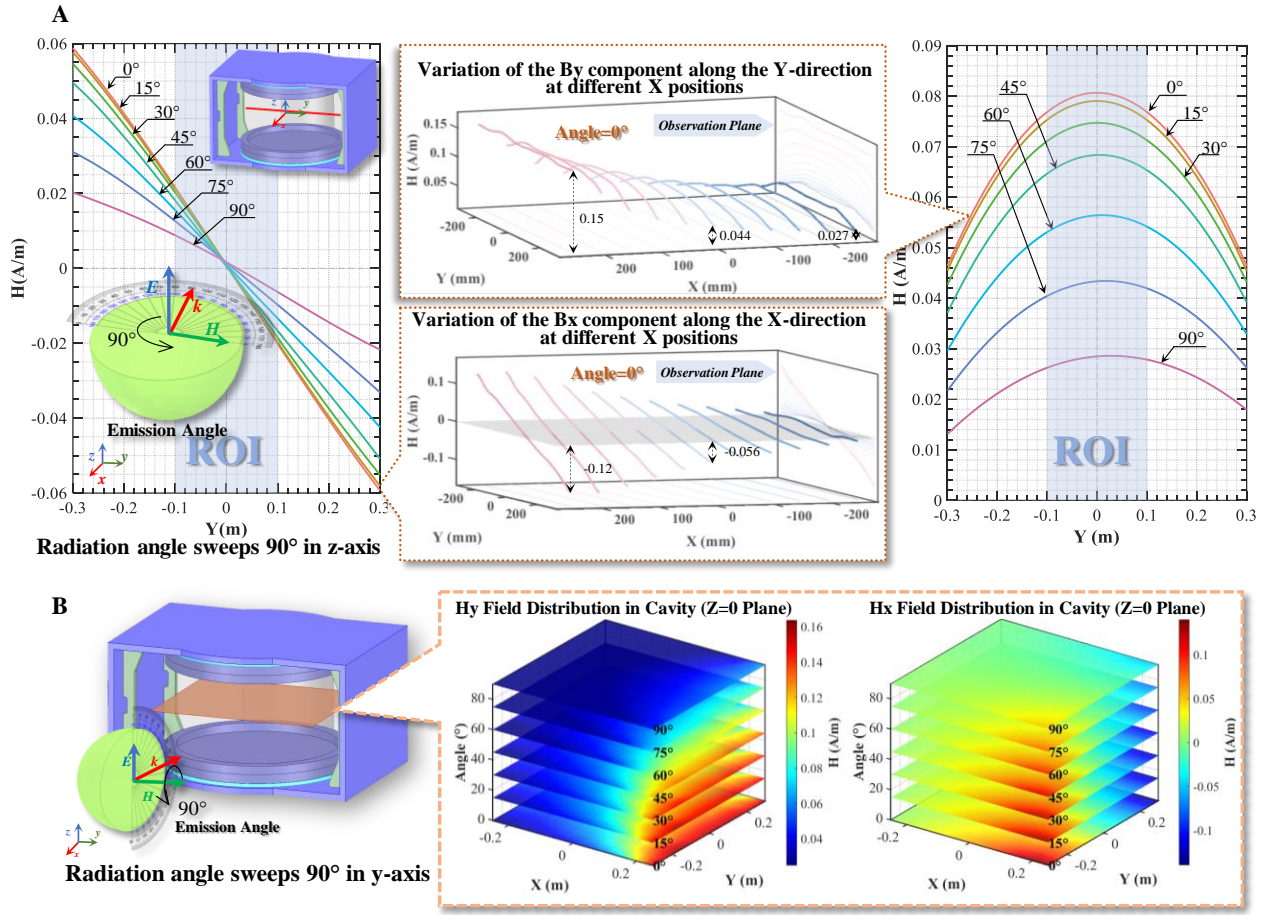

**Fig. 3. Effect of EMI incidence direction on magnetic field distribution in the cavity (origin at cavity center).** (**A**) Case I: a plane wave propagates along the −x direction with ***H*** ∥ ***y*** and ***E*** ∥ ***z***. With ***E*** fixed, the propagation vector ***k*** and magnetic field ***H*** were rotated around the ***E*** axis in 15° increments up to 90°. Shown are the longitudinal (*Hx*, left) and transverse (*Hy*, right) field distributions along the y-axis for different incidence angles. The central panel illustrates the *Hx* and *Hy* distributions at the Z=0 plane when the incidence angle is 0°. (**B**) Case II: with ***H*** fixed, the propagation vector ***k*** and electric field ***E*** were rotated around the ***H*** axis in 15° increments up to 90°. The resulting *Hy* (middle) and *Hx* (right) distributions at the Z=0 plane are shown for different incidence angles.

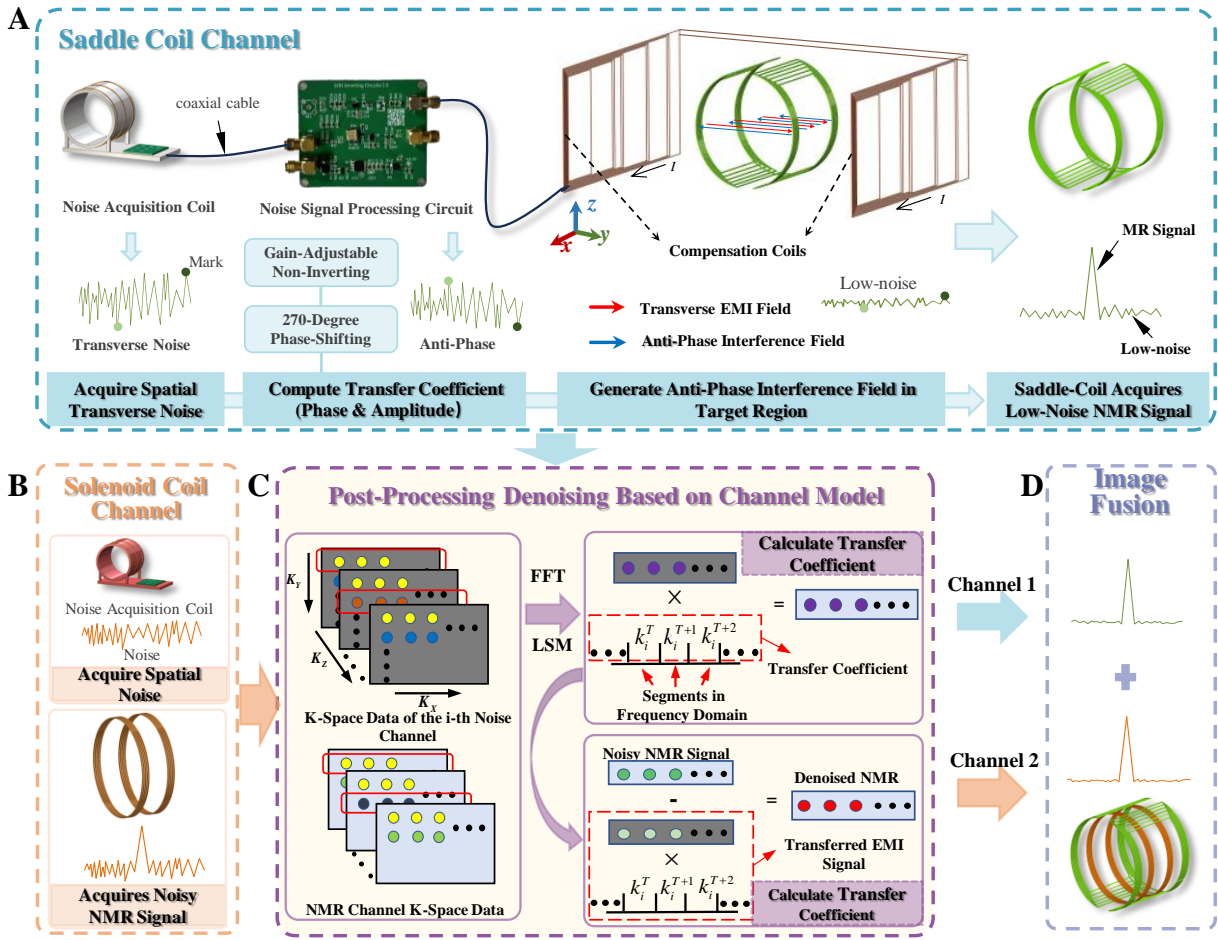

**Fig. 4. Active EMI Suppression Strategy** (**A**) Saddle-coil channel. The white coil denotes the spatial-domain detection coil, which monitors transverse EMI in real time. The detected signals are inverted by a processing circuit and fed to cancellation coils mounted on both sides of the cavity, generating an opposing field that suppresses transverse noise. The saddle coil then acquires low-noise MR signals under these conditions. (**B**) Solenoidal-coil channel. The red coil denotes a post-processing detection coil that acquires reference noise data for image denoising during reconstruction. The solenoidal coil directly acquires MR signals containing noise. (**C**) Post-processing model. A channel transfer-function approach is applied: peripheral k-space data (red boxes) are used to compute transfer coefficients across multiple frequency bands via least-squares fitting. These coefficients are then applied to the corresponding phase-encoding data to subtract the transferred EMI reference and obtain interference-free k-space data. (**D**) Fused output. The denoised MR images from both channels are combined to generate the final image.

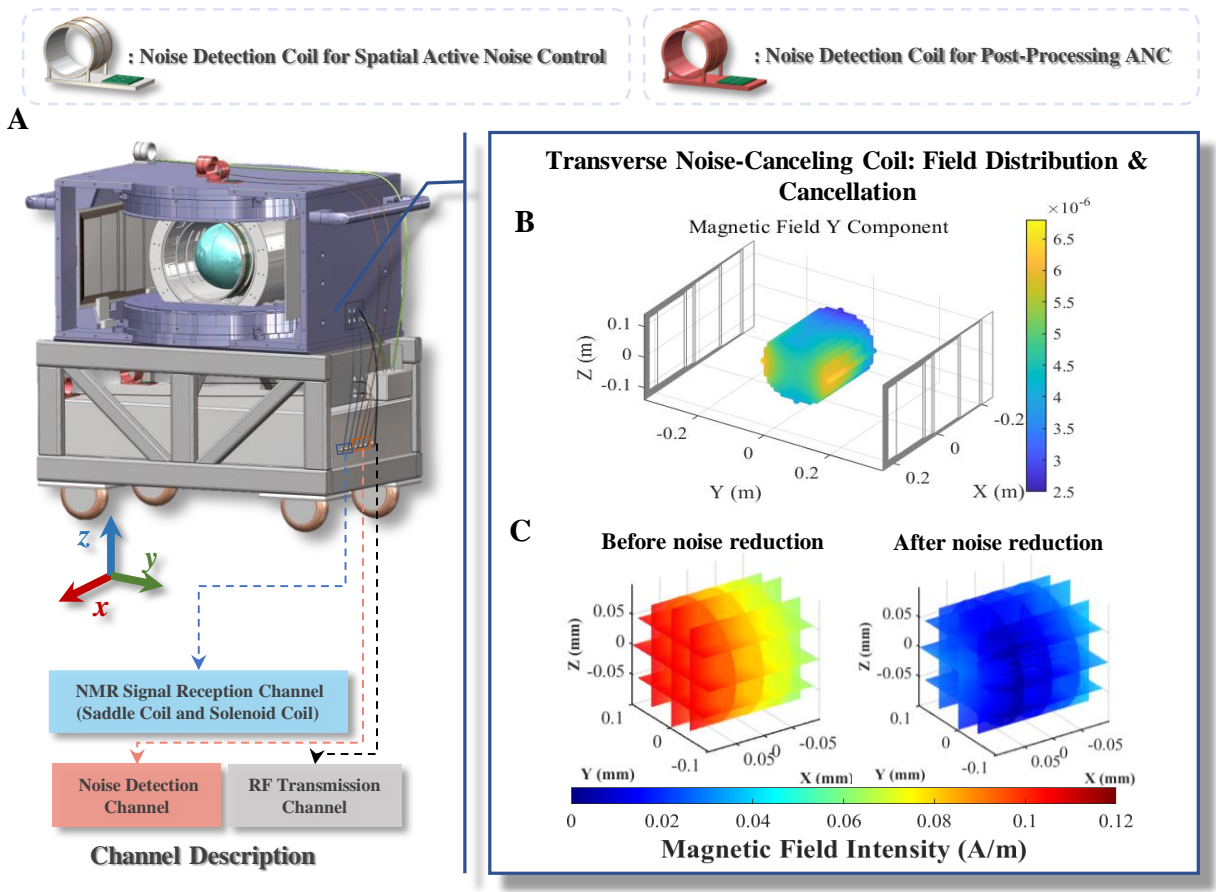

**Fig. 5. Installation and magnetic field distribution of the transverse cancellation coil.** **(A)** System configuration showing dual-channel orthogonal RF receive coils. White coils denote spatial-domain ANC detectors for real-time EMI monitoring, and red coils denote post-processing ANC detectors for noise modeling during image reconstruction. The spatial-domain ANC detectors are connected to a processing circuit (lower right), which inverts the detected signals and drives the transverse cancellation coils placed on both sides of the cavity. **(B)** Structure of the transverse cancellation coil, designed with a gradient-like winding to generate a strong field gradient along the y-axis. **(C)** Simulated performance of the spatial ANC scheme, demonstrating effective suppression of transverse interference in the imaging region.

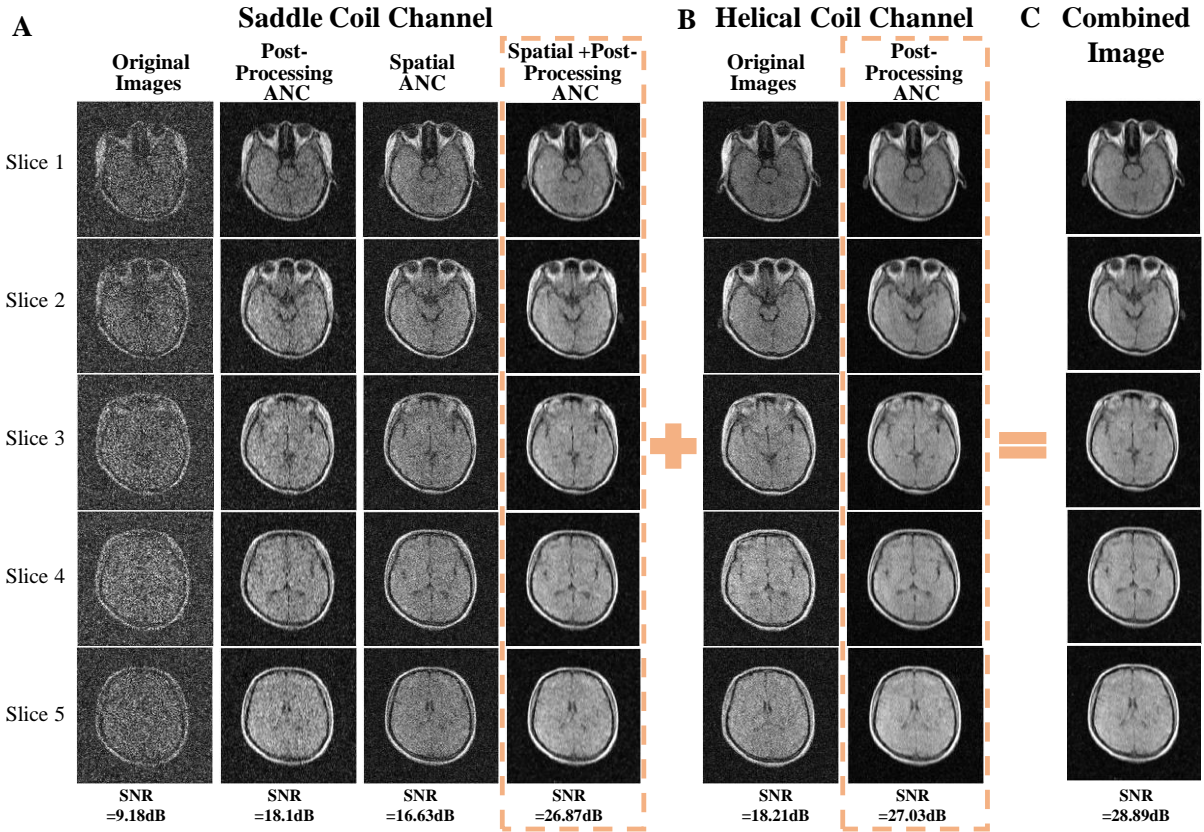

**Fig. 6. Dual-channel cranial imaging in an unshielded environment with active noise reduction.** (**A**) Sequential results from the saddle-coil channel: raw image, image after post-processing noise reduction, image with spatial-domain active cancellation, and image using the combined strategy. (**B**) Results from the solenoidal-coil channel: raw image and post-processed image, showing minimal degradation due to its lower sensitivity to transverse EMI. (**C**) Final fused image obtained by multi-channel fusion.

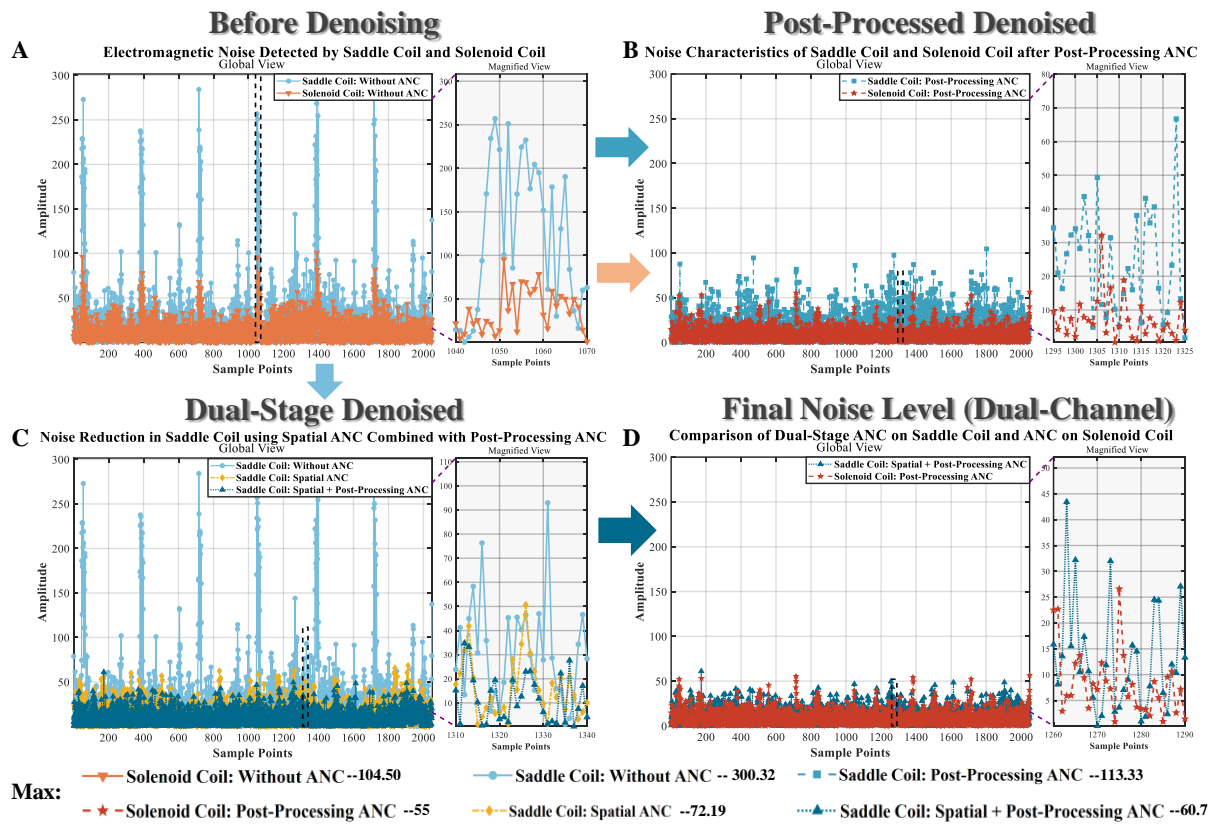

**Fig. 7. One-dimensional analysis of noise suppression under different strategies in an unshielded environment. (A)** Raw noise signals acquired by the saddle and solenoidal coils under ambient EMI. **(B)** Noise signals after applying the same post-processing reduction strategy to both coils. **(C)** Noise signals from the saddle coil with spatial-domain cancellation combined with post-processing (dual-mode suppression). **(D)** Comparison of the saddle coil after dual-mode suppression and the solenoidal coil after post-processing, showing convergence of noise levels.

**Funding:** This work was supported by:
    National Natural Science Foundation of China Grant 52077023 (X)
    National Natural Science Foundation of China Grant 52577249 (X)
    Chongqing Science and Health Joint Project under Grant 2023MSXM016 (X)
    Shenzhen Science and Technology Innovation Commission Grant CJGJZD20200617102402006 (X)
    Shenzhen Science and Technology Innovation Commission Grant KJZD20230923114110019 (W)

**Author contributions:**
    Conceptualization: JH, ZX, SS
    Methodology: JH, ZX, SS, JW
    Investigation: JH, YD, LT
    Visualization: JH, XK

Funding acquisition: ZX
Project administration: ZX
Supervision: ZX, JW
Writing—original draft: JH, SS
Writing—review & editing: JH, SS, ZX

**Competing interests:** The Authors declare that they have no competing interests.

**Data and materials availability:** All data are available in the main text or the supplementary materials.

# Supplementary Materials for

## Active noise cancellation in ultra-low field MRI: distinct strategies for different channels

Jiali He *et al.*

*Zheng Xu. Email: xuzheng@cqu.edu.cn

**This PDF file includes:**

Figs. S1 to S6

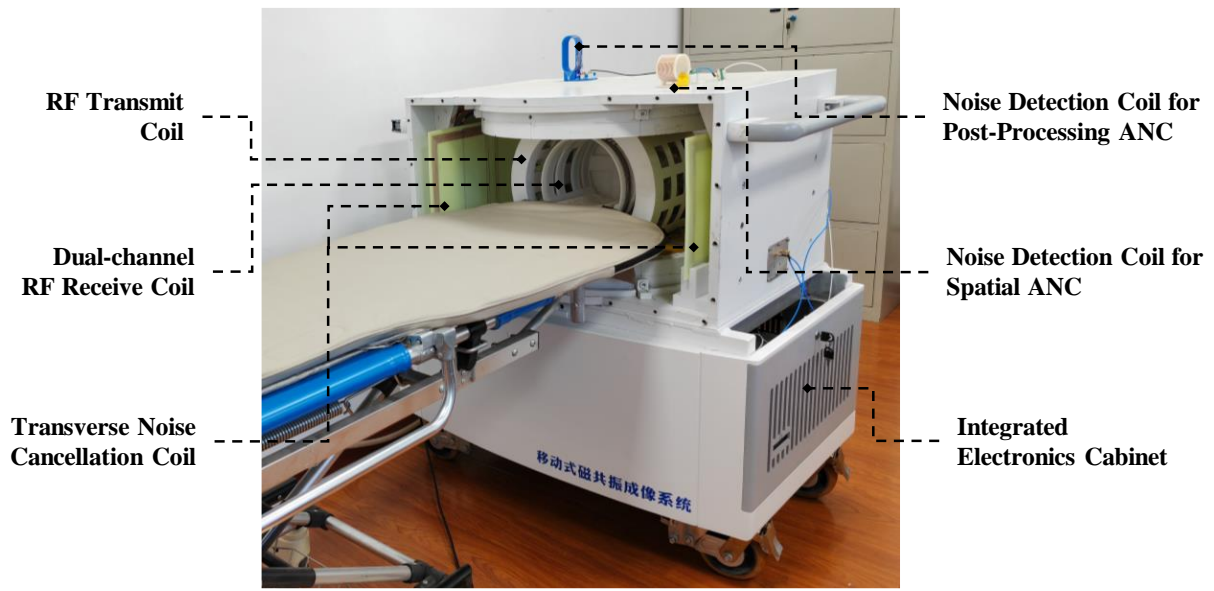

**Fig. S1. The internal structure of the ultra-low field MRI system.** A pair of transverse noise-cancellation coils is positioned on the left and right sides of the cavity. Two types of noise detection coils are arranged around the system (only the top coil is shown): one dedicated to post-processing active noise cancellation and the other to spatial active noise cancellation. The MRI receive setup employs a dual-channel orthogonal configuration combining a solenoid coil and a saddle coil.

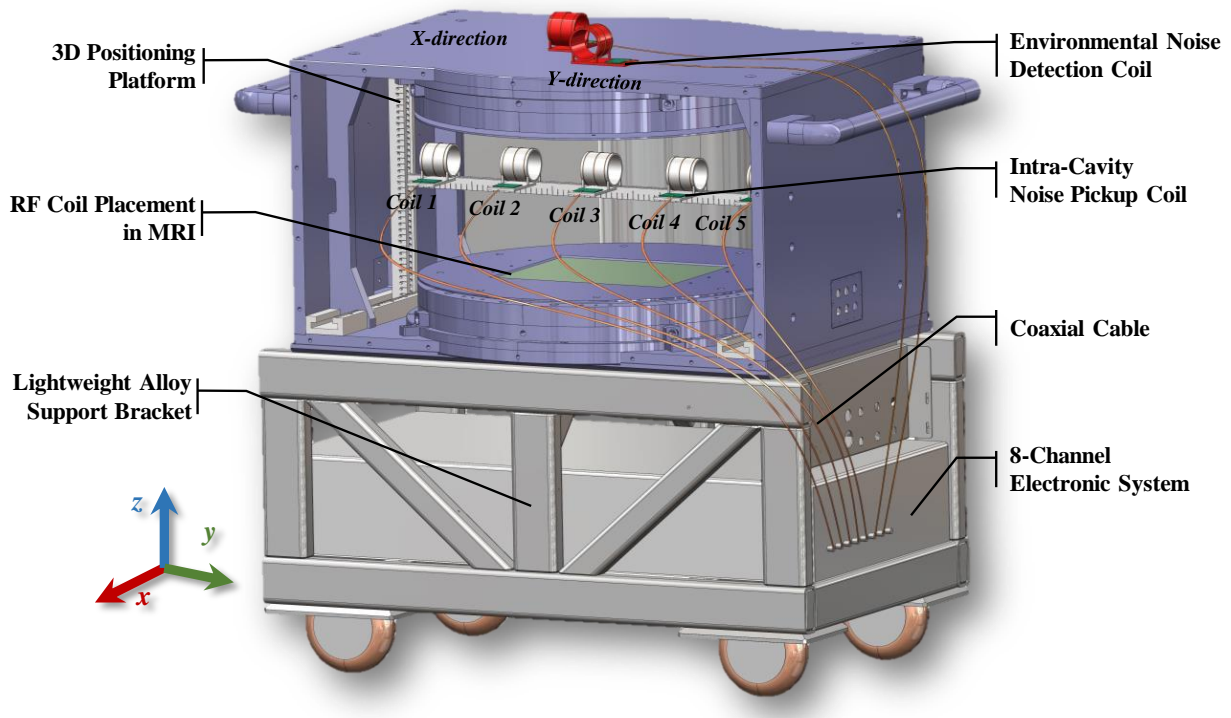

**Fig. S2. Electromagnetic interference (EMI) mapping system inside the cavity.** Five identical white detection coils are aligned along the Y-axis and mounted on a three-dimensional translation stage for spatial mapping of EMI within the imaging cavity. Two fixed red coils are positioned outside the cavity to monitor external EMI along the X- and Y-axes, providing reference data for modeling temporal variations of environmental noise during post-processing. All noise signals are acquired and processed using an in-house imaging system.

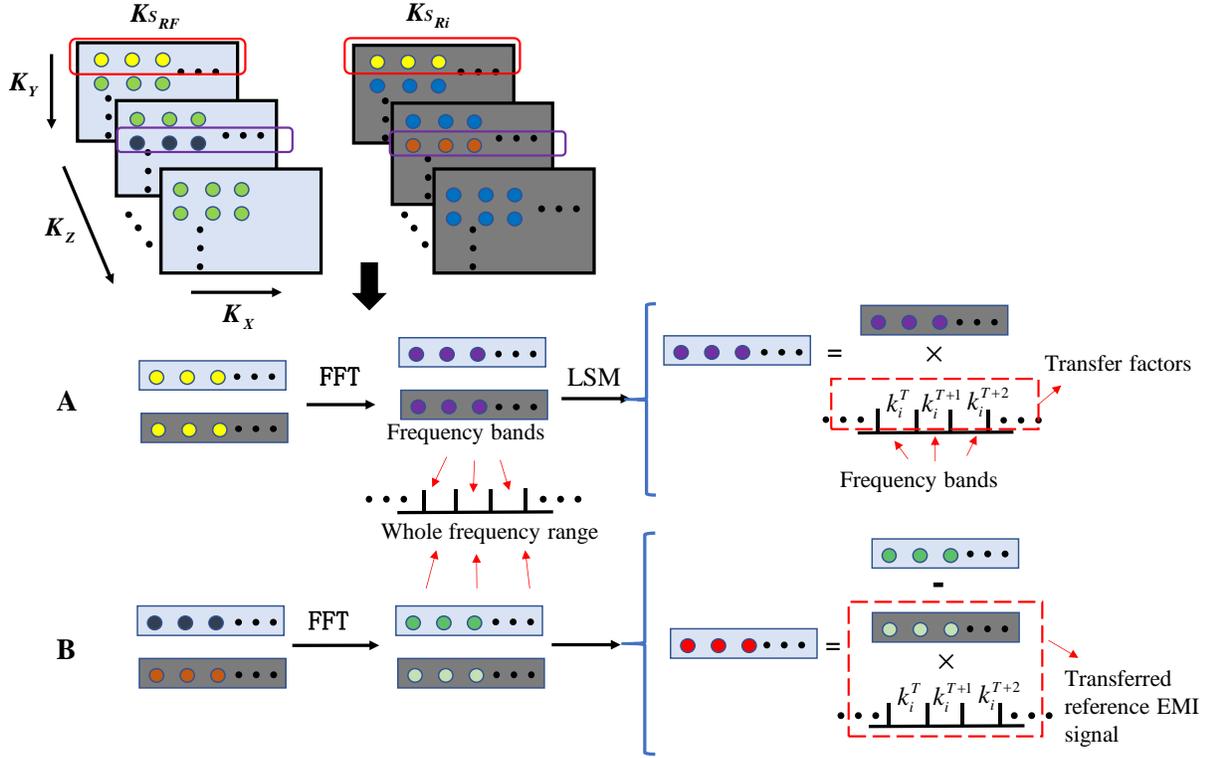

**Fig. S3. Diagram of active EMI suppression algorithm, which uses only one reference channel as an example.** Top: Two simultaneously acquired k-space data of the RF coil ($K_{SRF}$) and the ith EMI detector ($K_{SRi}$). The EMI detecting channels are termed as reference channels in this paper. The structure of the two k-spaces is the same, and their periphery data (in the red box) are chosen to calculate the transfer factors. The periphery data selected here are the first row in the first layer of both k-space data. (**A**) The periphery data in the frequency domain are decomposed into multiple frequency bands, and all bands cover the whole frequency range. The transfer factor in each frequency band is calculated using the least square method (LSM) on the basis of the periphery data in the same frequency band. (**B**) Then, other phase encoding data of the same location in the two k-spaces (in the purple box) are also decomposed into the same frequency bands. The interference-free k-space data of the RF coil could be obtained by subtracting the transferred reference EMI signals from the contaminated k-space data of the RF coil.

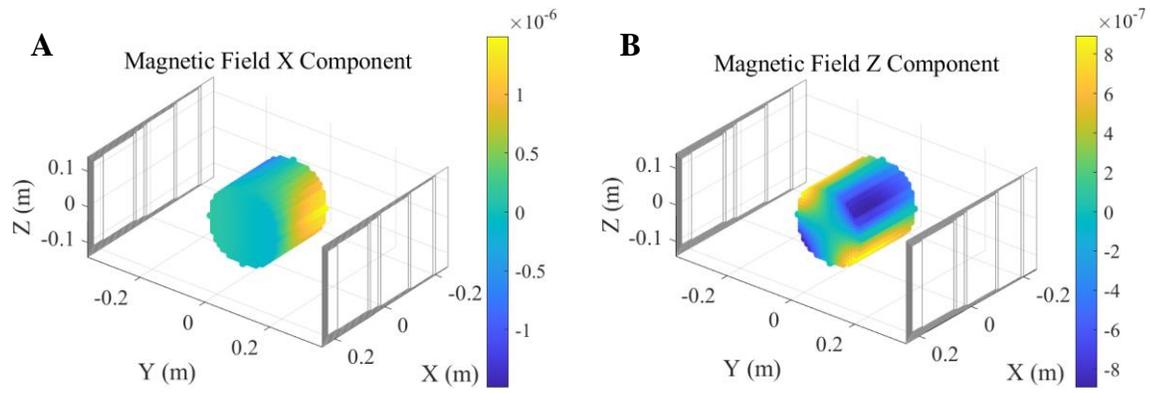

**Fig. S4. Magnetic field distribution generated by the transverse noise-cancellation coils in the target region.** (**A**) The X-component exhibits a symmetric positive–negative pattern with small amplitude, resulting in near-zero net magnetic flux within the solenoid receive coil and thus avoiding additional interference in the solenoid channel. (**B**) The Z-component is negligible and does not introduce significant artifacts in MRI signal acquisition.

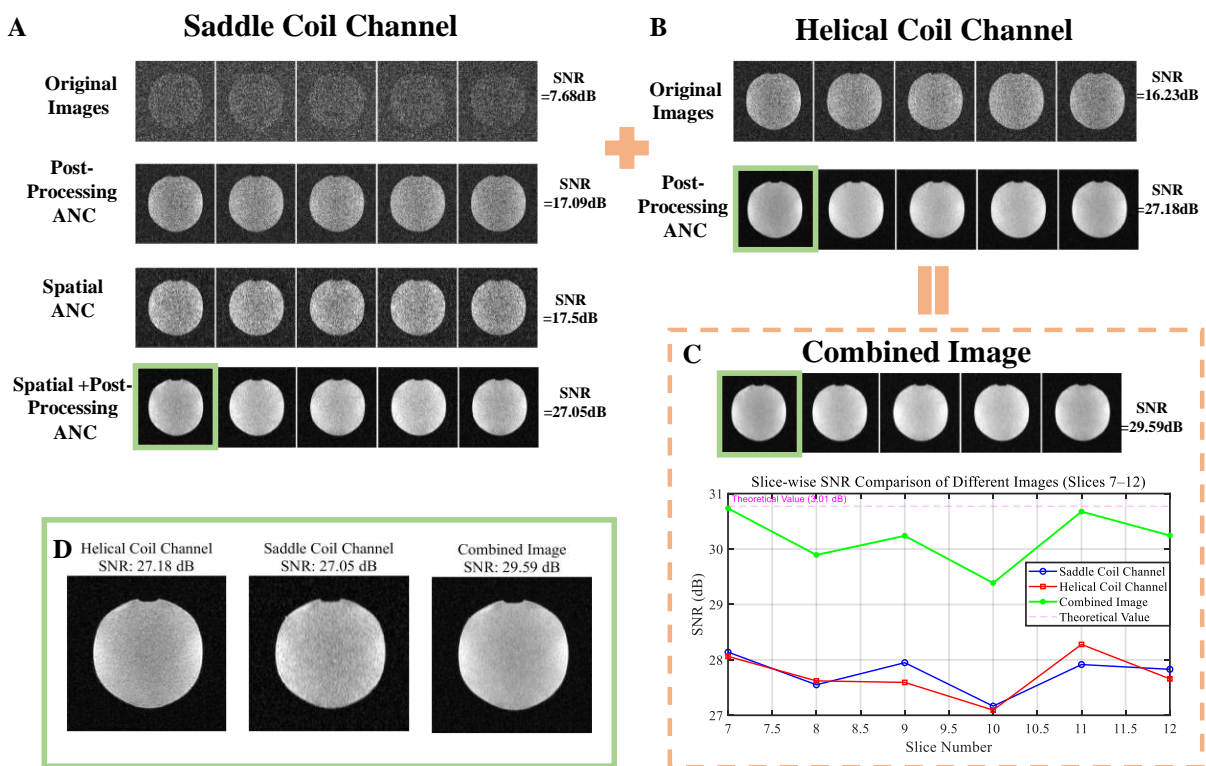

**Fig. S5. Dual-channel water phantom imaging in an unshielded environment with different noise-cancellation strategies. (A)** Sequential results from the saddle channel: raw image, image after post-processing cancellation, image after spatial-domain cancellation, and image after the combined strategy. **(B)** Results from the solenoidal channel: raw image and post-processed image. **(C)** Fused dual-channel images and SNR across slices. **(D)** Enlarged views of selected regions from individual channels and fused images.

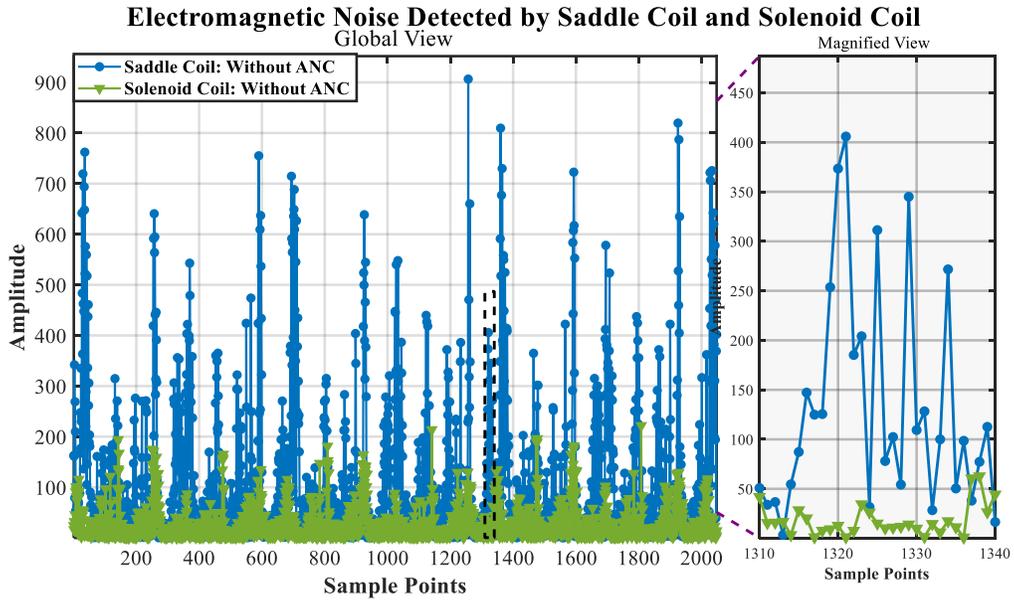

**Fig. S6. One-dimensional noise comparison between the solenoid and saddle channels before noise cancellation.** Under stronger interference conditions, the noise amplitude detected by the saddle coil is approximately 4.5 times higher than that detected by the solenoid coil.